# Chapter 1. Autonomous Intelligent Cyber-defense Agent: Introduction and Overview

Alexander Kott

## Abstract

This chapter introduces the concept of Autonomous Intelligent Cyber-defense Agents (AICAs), and briefly explains the importance of this field and the motivation for its emergence. AICA is a software agent that resides on a system, and is responsible for defending the system from cyber compromises and enabling the response and recovery of the system, usually autonomously. The autonomy of the agent is a necessity because of the growing scarcity of human cyber-experts who could defend systems, either remotely or onsite, and because sophisticated malware could degrade or spoof the communications of a system that uses a remote monitoring center. An AICA Reference Architecture has been proposed and defines five main functions: (1) sensing and world state identification, (2) planning and action selection, (3) collaboration and negotiation, (4) action execution and (5) learning and knowledge improvement. The chapter reviews the details of AICA's environment, functions and operations. As AICA is intended to make changes within its environment, there is a risk that an agent's action could harm a friendly computer. This risk must be balanced against the losses that could occur if the agent does not act. The chapter discusses means by which this risk can be managed and how AICA's design features could help build trust among its users.

## 1. Introduction

This book is based on the premise that the future of cyber-defense and cyber resilience will depend largely on autonomous, artificially intelligent (AI) agents. Such an agent will reside on a system that includes one or more computing devices and be responsible for defending the system from cyber compromises. If a compromise occurs, the agent will then be responsible for response and recovery of the system, usually autonomously. To refer to such a class of agents, we use the term Autonomous Intelligent Cyber-defense Agent (AICA). In this book, we explore how AICA will be designed and how it will operate.

Experience shows that even well-protected computing systems are likely to be successfully attacked and infiltrated by hostile malware. There is no reason to believe this will be any different in the future. Today, when a compromise occurs, response, mitigation and recovery depend largely on human cyber-defenders. This approach is becoming increasingly untenable. With an ever-growing number of computerized, automated and even autonomous systems in our society, human-based cyber-defense must be replaced by autonomous cyber-defenders such as AICA.

Similarly to the current generation of cyber-defense tools, AICA will detect malicious signatures, patterns and anomalies; it will also classify, characterize and diagnose what it observes within its environment, traffic and host. However, unlike the current generation of cyber-defense tools, AICA is a doer, not merely a watcher. It will have to plan and then decisively execute responses to attacks and perform recovery actions.

AICA will be an active fighter in maintaining a system's resilience (Kott and Linkov 2019) against cyber threats. This means that the agent's capabilities should include a significant degree of autonomy and intelligence for the purposes of rapid response to a compromise – either incipient or already successful – and rapid recovery that aids the resilience of the overall system. Often, the response and recovery efforts need to be undertaken in absence of any human involvement, and with intelligent consideration of the risks and ramifications of such efforts.

The cyber-defense technology community is beginning to recognize the potential and even necessity of autonomous, AI-supported cyber-defenses. In particular, the vision of AICA is a product of a NATO-based research project that took place from 2016-2020. The research yielded an AICA Reference Architecture (Kott et al. 2018). Later, an international working group formed to continue work on AICA (see https://www.aica-iwg.org/). The authors of this book are grateful to the AICA research community.

In the remainder of this chapter, we discuss what it means for AICA to be an agent, what environments that agent will face, what roles the agent will perform, how these roles will be supported by the internal architecture of the agent, and the inevitable concerns regarding the risks and trust associated with such an autonomous technology. The chapter concludes with a preview of each chapter of the book.

## 2. AICA as an Agent

We call AICA an agent. What does it mean? The term "agent" refers to software or collection of software that resides and operates on one or more computing devices, perceives and comprehends its environment, and plans and executes purposeful actions on the environment (and itself) to achieve the agent's goals.

Autonomy, complete or partial, is an important characteristic of an agent. AICA will have to be capable of autonomous planning and execution of complex, multi-step activities. These activities will pursue the key goal of the agent – defeating or degrading malware while anticipating and minimizing any resulting side effects. It will have to be capable of adversarial reasoning (Kott and McEneaney 2006) to battle against thinking, adaptive malware. To defend itself against the malware, AICA should keep itself and its actions as undetectable as possible, and thus will have to use deceptions and camouflage.

The autonomy of such an agent is a necessity, not luxury. Today, much of cyber-defense is based on remote monitoring and remote mitigation and recovery (Kott and Arnold 2013). However, today's reliance on human cyber-defenders, whether local or remote, will be untenable in the future for a number of reasons.

One reason is the growing scarcity of human cyber-experts to defend systems, either remotely or onsite. This is further exacerbated by the proliferation of robots, such as self-driving cars, where a local cyber-defender is unlikely to be found by definition – these systems are intended to operate with little or no human involvement (Kott and Stump 2019). There is an ever-growing number of critically needed software systems, all of which present tempting targets for malicious cyber actors and require cyber-defenders in ever-growing – unsupportable – quantities.

Remote cyber-defense, i.e., remote monitoring, mitigation and recovery, is further complicated by the growing sophistication of malware. One of the first actions sophisticated malware may do is degrade or spoof the communications of a system that uses a remote monitoring center. This means that the system must possess local defenses that do not depend on communicating with the remote service. If a local human defender is not available, as will most often be the case, the system will need to rely on a local autonomous agent, AICA.

Let's take a more detailed look at the features and characteristics that should be exhibited by any autonomous intelligent agent (Théron et al. 2019). These are helpful as we consider the comparable features and characteristics of AICA.

A proper agent should be assigned a specific mission, with corresponding goals and constraints. It should possess the key competencies to execute that mission, such as the ability to perceive the environment in which the agent is deployed, detect attacks, plan and assess the required countermeasures, and adapt rapidly to successes or failures when executing its plan.

The agent should be proactive and autonomous, which means it should not rely on an external source to initiate or control its activity. On its own, the agent should be able to assess the situation, and make decisions and execute actions, without being controlled by another program or a human operator. To do so, the agent typically needs a base of knowledge, memories of what the agent has done before, and in many cases, even a degree of self-learning from experience.

Safety is another important characteristic. The agent should not harm the friendly systems it defends. For that, the agent should be able to anticipate the ramifications of its actions and attempt to minimize the risk of causing harm. In exceptional cases, considerations of safety may require the agent to contact a remote human controller, activate a fail-safe mode or even self-destruct when no other possibility is available. Similarly, the agent should be trustworthy, e.g., it will not deceive other friendly agents or human operators.

Finally, let's consider robustness and resilience to various threats and abnormal circumstances. Doing this requires the agent to possess a means of defending itself and recovering its own operations when degraded by a threat.

To achieve all of these characteristics, should an agent be a monolithic piece of software? That can be one implementation option. In general, however, an agent's modules should be distributed over multiple processes or devices, or implemented as a team of agents or subagents.

If implemented as a multi-agent system, a number of additional considerations must be addressed. These include the manner in which the multiple, potentially heterogenous agents are coordinated, self-organize, admit (or not) new members and deal with emergent behaviors.

## 3. AICA's Environment

AICA operates within its environment, i.e., everything that surrounds the agent and that the agent can perceive. This includes the computer hardware and software where the agent operates; the physical entity controlled by the computers, e.g., a self-driving car; the malware; the humans who communicate with the agent or with surrounding hardware and software; and other agents that the agent can find and with whom it can communicate.

To make our discussion more concrete and focused, let's consider a single physical item or platform, such as a vehicle or industrial robot, with one or more computers residing on the platform connected to sensors and actuators. We assume that, at any given time, one or more computers are compromised by malware. The compromise is either established as a fact or is suspected. We further assume that, in general, the platform's communications with any remote operators or a monitoring center is compromised as well; malware has disabled or is spoofing the communications.

As mentioned earlier, with compromised communications, conventional centralized cyber-defense is often infeasible. Here, by "conventional centralized cyber-defense," we mean an architecture where local sensors send cyber-relevant information to a central location where highly capable cyber-defense systems and human analysts detect the presence of malware and initiate corrective actions remotely. It is unrealistic to expect that the human cyber-defenders will reside on the platform, for example, a self-driving vehicle, or that they would have the necessary skills or time available to perform cyber-defense functions locally on the vehicle even if present.

Furthermore, many situations demand much faster responses than human responders may be able to provide. Criminals or irresponsible pranksters are able to take control of cars traveling at high speed or planes in the air, which may constitute a mortal threat to the vehicle's passengers and others interacting with those systems. In such cases, waiting for a human incident response team will not do. Instead, such systems need an onboard intelligent autonomous agent capable of taking the necessary response and recovery actions, with response times on the order of seconds or even less (Kott and Théron 2020). In short, AICA operates in an environment where it must act autonomously.

But how did AICA find itself in this environment? We assume here that the agent resides on a computer where it was originally installed by a human controller or an authorized process. We do envision a possibility that an agent may move itself (or a replica of itself) to another computer. Such propagation is assumed to occur only under exceptional and well-specified conditions, and takes place only within a friendly network – from one friendly computer to another friendly computer. Granted, this type of action might seem very close to the controversial idea of a "good virus" (Muttik 2016). However, AICA is not a virus, because it does not propagate except under explicit conditions within authorized and cooperative nodes.

## 4. AICA's Roles

Having considered the demanding environment of AICA, let's explore its roles within that environment.

Unlike most of today's cyber-defense tools, AICA is a doer, not merely a watcher (Kott and Théron 2020). Most of today's tools are largely watchers: they monitor traffic and events; check packets and files; detect malicious signatures, patterns and anomalies; and classify and characterize what they watch. In some respects, such tools can also be classified as doers: they issue alerts, stop suspicious packets and connections, and remove or quarantine suspected malware.

Still, such tools are very constrained and limited in what they do. In the face of a sophisticated and ongoing attack by a capable, stealthy malware, today's tools do little to plan, assess options, and execute a sophisticated, multi-step response.

Further, when malware succeeds in degrading a friendly system, today's tools do little to plan and execute recovery activities. The critical activities of response and recovery after a successful cyber-attack are left to the human cyber analysts, incident responders and system administrators. As just discussed, these human actors are unlikely to be available in the environments where AICA operates. AICA has to perform these activities, and do so autonomously.

Granted, AICA has to be a competent watcher too. The agent must be able to observe the state and activities within the system it is asked to defend. Using these observations, AICA must be able to diagnose the situation, understand what is happening and project the future, i.e., the likely actions of the malware and how those actions would affect the state of the system.

Having assessed the situation and formed a vision of anticipated future states if the malware is unopposed, AICA must create a plan of action, or generally, several alternative plans. All such plans have a degree of uncertainty, and AICA should anticipate possible adaptations of its plans as well.

With one or more plans available, the agent should assess the risks and benefits involved (Kott et al. 2017) and make its decisions accordingly. Needless to say, all this deliberation must be performed very rapidly. In cases when time is lacking and immediate action is needed, instead of engaging in such reasoning, AICA may have to resort to simple but fast "condition-action" rules.

Once a plan is selected, AICA executes the actions. Some of the actions might be benign, e.g., gathering additional information. Other actions, however, may have destructive impacts, such as destroying, degrading or quarantining certain software and data – autonomously – or inhibiting certain actions of the malware. This may involve stopping or starting certain processes, installing or reinstalling software, or initiating or terminating connections. In executing such actions, AICA must observe the specified rules of engagement and continually assess risks (Ligo et al. 2021).

One of the major risks facing AICA is being destroyed by the enemy malware. In the case where the enemy malware knows that an agent is likely to be present on the computer, the malware will seek to find and destroy the agent. Therefore, a key responsibility of AICA is self-defense and self-preservation. The agent must possess techniques and mechanisms for maintaining a certain degree of stealth, camouflage and concealment. More generally, the agent must take measures that reduce the probability that the enemy malware would detect it.

On the other hand, AICA may find it advantageous to communicate with other friendly agents that might reside on other computers and systems. For example, AICA may need to ask another agent to terminate a certain connection or send software to AICA. Such a collaboration entails risks because it potentially reveals the presence and activities of AICA to the malware. In cases where the communications may be impaired or observed by the malware, the agent may have to eschew collaboration and operate alone. Nevertheless, in general, the agent should be able to collaborate with other friendly agents when a need arises and conditions permit. Thus, collaboration schemes and negotiation mechanisms are needed for that.

Finally, a friendly agent with a particularly important role is the human operator. Typically, these operators would be the personnel of a remote operations center who deploy, monitor and control AICA, to the extent that communication channels permit. The agent, whenever requested and when conditions permit, reports its situation, activities and related data to the external controller. This information helps the controller to make inferences about the trustworthiness of the agent, measure the effectiveness of the agent (Kott and Linkov 2021) and determine whether AICA needs to receive updates.

More generally, we envision a degree of role distribution between AICA and a remote cyber-defense control center (Kott et al. 2021). Their roles are not incompatible and may coexist. Both have their strong and weak points. As discussed, relying primarily on remote monitoring and response may be risky or impossible if the sophisticated malware takes over the communication channels. If remote mitigation of a cyber compromise cannot be provided rapidly, the compromised system will find itself at extreme risk. This is when AICA is necessary, even if it is less capable then the comprehensive cyber-defense capabilities of a competent remote center.

Similarly, reliance only on AICA comes with its own risks. For example, if the malware is able to overcome the capabilities – inevitably limited – of AICA, external intervention will be a necessity. Such coexistence should be carefully orchestrated. In particular, a clear protocol should be established for the handover of responsibilities between AICA and the remote center, and back. If both need to operate simultaneously, a coordination protocol should ensure that their respective actions do not produce undesirable interference.

## 5. AICA's Architecture

(This section is based in part on a book chapter. Théron, P., Kott, A., Drašar, M., Rzadca, K., LeBlanc, B., Pihelgas, M., Mancini, L. and de Gaspari, F., 2020. Reference architecture of an autonomous agent for cyber defense of complex military systems. in adaptive autonomous secure cyber systems (pp. 1-21). Springer, Cham.)

The sophisticated roles and responsibilities of AICA demand appropriate internal functions. Let's consider the functional capabilities the agent must possess within its architecture (Kott et al. 2018; Théron et al. 2020). Fig. 1 depicts the functional components of the agent.

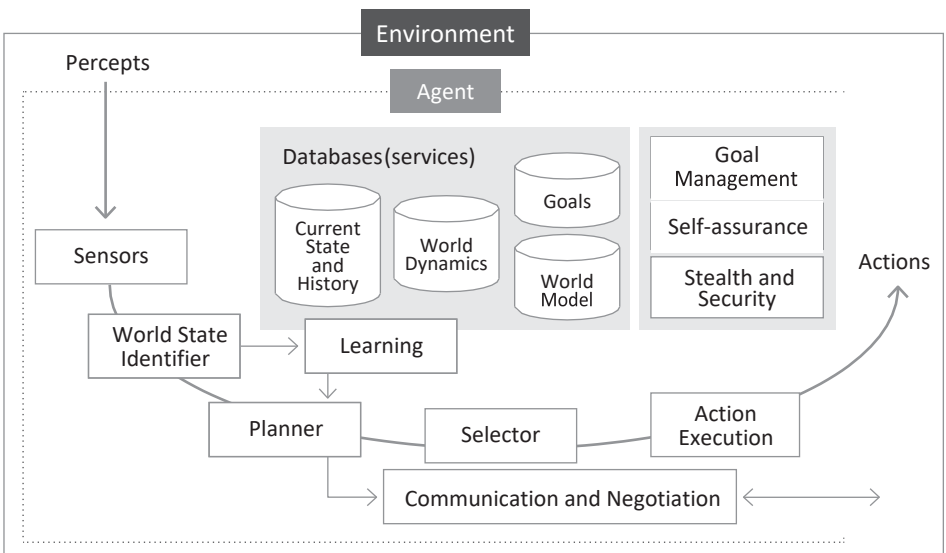

Fig. 1 Functional components of an agent.

The AICA Reference Architecture (Kott et al. 2018; Théron et al. 2020) defines five main functions:

- Sensing and world state identification
- Planning and action selection

- Collaboration and negotiation
- Action execution
- Learning and knowledge improvement

Sensing and world state identification allows a cyber-defense agent to acquire data from the environment in which it operates as well as from itself in order to understand the current state of the world. This sensing and world state identification function relies upon the "world model," "current world state and history," "sensors" and "world state identifier" components of the functional architecture. Current world state descriptors are captured by the agent's sensing function, while the world state identification draws from (1) processed current world state descriptors and (2) learned world state patterns. Having identified a problematic current world state pattern (e.g., a cyber compromise), the world state identification function triggers the planning and action selection function.

The planning and action selection function allows AICA to formulate one or several action proposals and then submit them to the action selector. The latter decides on the action or a set of actions to execute in order to resolve the problematic world state pattern previously identified by the world state identifier function. This function relies upon the "world dynamics" and should include knowledge about the "actions and effects," "goals," "planner-predictor" and "action selector" components of the functional architecture. The planning function operates based on (1) the problematic current world state pattern and (2) a repertoire of response actions. The action selector function operates based on (1) the proposed response plans, (2) the agent's goals and (3) execution constraints and requirements, such as the environment's technical configuration. The proposed response plan is analyzed by the action selector in the light of the agent's current goals as well as any execution constraints and requirements. The proposed response plan is then trimmed of elements that do not fit the situation at hand and augmented with prerequisite, preparatory, precautionary or post-execution complementary actions. The action selector thus produces an executable response plan, which is then submitted to action execution, after collaboration and negotiation if needed.

The collaboration and negotiation function enables AICA to (1) exchange information with other agents or a central cyber command and control (C2), for instance, when one of the agent's functions cannot reach satisfactory conclusions on its own and (2) negotiate with its partners the details of a consolidated conclusion or result. Collaboration and negotiation operate based on (1) the outgoing data sets sent to other agents or to a central C2, (2) incoming data sets received from other agents or a central cyber C2, and (3) the agents' own knowledge (i.e., produced through its function of learning and knowledge improvement). When an agent (including possibly a central cyber C2) develops conclusions, it shares them with other (selected) agents, usually including the one that issued the original request for collaboration. Once this response is received, the network of involved agents starts negotiating to develop a consistent, satisfactory set of conclusions.

The action execution function enables AICA, based on the action selector's executable response plan, to monitor its execution and its effects, and provides the means to adjust the execution of the plan (or possibly dynamically adjust the plan) as needed. This function relies upon the "goals" and "action execution" components of the functional architecture. Action execution includes four subfunctions: (1) action effector, (2) execution monitoring, (3) effects monitoring and (4) execution adjustment. Taking into account the environment's technical configuration, the action effector function executes each planned action within the executable response plan in the scheduled order. The execution monitoring function uses the executable response plan in concert with plan execution feedback to monitor each action's execution status. Any status apart from "done" triggers the execution adjustment function. The effects monitoring function operates on the basis of the executable response plan and environment's change feedback, leading to execution adjustment as needed. Should warning signs be identified by one of the two previous functions, the execution adjustment function will either adapt the actions' implementation to circumstances or modify the plan.

The learning and knowledge improvement function allows AICA to use its experience in order to improve progressively its efficiency. This function relies upon the learning and goals modification components of the functional architecture. The learning function operates based (1) feedback data from the agent's functions and (2) feedback data from the agent's actions. This function analyzes the reward function of the agent (the distance between goals and achievements) and their impact on the agent's knowledge database. Results are fed to the knowledge improvement function. The knowledge improvement function merges the results (propositions) from the learning function with the current elements of the agent's knowledge.

Like any reference architecture, this proposed AICA Reference Architecture is merely a step toward a structured solution and a set of common vocabulary with which to discuss possible implementations. Actual implementations of AICA may differ dramatically. Several chapters in this book describe case studies of implementations of AICA-like agents and illustrate the diversity of possible approaches.

## 6. AICA's Risk and Trust

The architecture discussed in the previous section prompts at least two observations. One is that it is necessarily complex. In a complex software system, much can go wrong and ensuring highly reliable operation of such a system is difficult and expensive. Second, the entire architecture is aimed at actively doing things, i.e., making changes in AICA's environment. We already mentioned that in order to fight the malware that has infiltrated the friendly computer, the agent may have to take destructive actions, such as deleting or quarantining certain software. Granted, such destructive actions should be carefully controlled by the appropriate rules of engagement and are allowed only on the computer where the agent resides. Needless to say, developers of AICA will strive to design its actions and planning capability to minimize the risk to the system (Ligo et al. 2021).

Nevertheless, in general, such risk cannot be fully eliminated. Nothing can guarantee that AICA will always preserve the availability or integrity of the functions and data of the computer the agent is trying to defend. It is not entirely improbable that the agent will "break" the friendly computer, disable important friendly software, or corrupt or delete important data. AICA's actions may have harmful consequences of a functional, safety, security or ethical nature.

To be sure, this is nothing new. Every technology comes with risk. Any artifact may cause unintended harm. The reason to accept AICA, as with any technology, is for the users to determine that the advantages of using the technology outweigh the risk that comes with it. In case of AICA the risk that the agent's action will harm a friendly computer must be balanced against the losses that might occur if the agent does not act.

This is a not fully comforting answer. Can we do better than that? Can we find other ways to manage the risk? Well, a natural reaction to a risky machine is to have a human supervise the machine. Perhaps, we should create a human-AICA team, where the human could intervene in AICA's operations as needed? Unfortunately, this is unlikely to produce positive outcomes.

Let's consider an analogy. Given AICA is an autonomous agent, a suitable analogy is an autonomous, self-driving car. When a dangerous situation arises on a road, e.g., a pedestrian suddenly appears in the middle of a street, should the human passenger take over the controls and try to swerve around? Or should the human let the autonomous driving system execute its collision-avoidance routine? The answer probably depends on who is more likely to avoid the collision – the human or autonomous agent. Chances are quite high that the human – who has been driving less since buying the autonomous car – is not as alert or capable as the autonomous driving system. If so, the best course of action (COA) is for the human not interfere with the agent's driving.

The case of cyber-defense is even more unfavorable for the human. Car driving, after all, was initially specifically designed for human drivers. It has become relatively natural for humans to drive a car. Many are quite good at driving. In the world of cyber-defense, little if anything was designed for humans. The extremely high volume of information, the extremely short durations of events and so on are inconsistent with human cognitive abilities. Thus, the chances of a human successfully interfering with an autonomous decision of AICA are even lower than in case of a human driver taking over the controls of an autonomous car.

Still, other ways exist for a human to influence AICA. At the stage of AICA's software development, human designers determine the goals of the agent and populate the knowledge base of the agent. They decide on decision algorithms and decision criteria. If AICA's knowledge base is formed through machine learning, human designers select the data samples for training and guide the learning process. At the validation stage, human developers create an ensemble of test cases, assess the correctness of the agent's behavior and measure its effectiveness (Ligo et al. 2021). In these multiple ways, humans shape the behavior of AICA before it begins its actual operation.

Then, once AICA is placed in operation, human supervisors can observe the agent's behavior and determine whether its behavior meets the desired criteria. If the behavior needs adjustments, a human supervisor can take action. As already discussed, it would be unwise to intervene into the agent's fast-paced operation directly. Instead, the human supervisor, in a deliberate fashion, can modify the goals, criteria and constraints of AICA, or offer additional examples for the agent's learning process. All this can be done without taking AICA offline, while it continues its operations.

Trust is closely related to risk. Whenever a technological artifact is perceived as associated with risk, human users have difficulties trusting the artifact. Undoubtedly, human users will build their trust of AICA only gradually, by observing its behavior over a period of time, in multiple events. As they observe the agent's behavior, they will interpret its behavior, i.e., try to determine what exactly AICA did and for what reasons. Eventually, the human users will accumulate enough evidence that AICA appears to do the right things, for the right reasons. AICA's designers can help this trust-building process by providing the agent a means to communicate to the human users what the agent is doing, the decision process involved, and the information used as inputs into its decision and eventual actions (Linkov et al. 2020).

## 7. Preview of this Book

The next chapter "Alternative Architectural Approaches" describes an approach to AICA and considerations about the rationale of the design of AICA's architecture. Further, the Multi Agent System Centric AICA Reference Architecture (MASCARA) is presented with regard to the three layers of its definition: general, detailed and technical. From the early prototyping experience, lessons for the future are drawn.

Next, the "Perception of Environment" chapter describes how AICA continually perceives (obtains information about) its environments (network, host computer(s) hardware and software, broader systems such as a vehicle on which AICA resides, etc.). Perception in AICA is best considered a pipeline consisting of four main parts: physical sensors, logical sensors, transformers and the world state. This chapter addresses the complexity surrounding its perception, providing guidelines and state-of-the-art examples.

Because AICA exists to fight against a cyber-adversary, it needs a means to perceive / sense the presence and characteristics of hostile agents (malware) and their actions and effects, as well as recognize when appropriate active sensing may be engaged. The "Perception of Threat" chapter discusses key use cases and methods of obtaining threat intelligence, fingerprinting, characterizing threats, passive threat detection, anomalous activity detection, the use of honeypots and threat hunting.

Given the perception of its environments and threats, AICA attempts to assess and characterize the situation. In "Situational Understanding and Diagnostics," we discuss situational understanding (SU) inputs from sensors, the dependencies of SU on the knowledge base, updating the knowledge base as needed for SU, and logical formalisms supporting knowledge and reasoning. We also discuss the means of using abstraction and generalization, through which agents can better manage model complexity, using illustrative examples.

The "Learning about the Adversary" chapter considers how an agent can gain insights about the behaviors and intents of the adversary (human-directed attack or automated malware). We argue that the evolving nature of cyber-adversary tactics and techniques and system configurations and vulnerabilities makes it difficult for autonomous agents to rely on supervised learning or, in general, much a priori expert knowledge. We review alternative approaches and recent advancements in the field.

Having examined the perception of the environment and threat, and the overall assessment of the situation, in "Response Planning," we discuss how AICA plans a COA, or multiple COAs, intended to defeat the malware and/or minimize the malware's damage to the system. We explore ways to include an adversary model into the defender's decision making, to understand how observations differ from the known adversary model and hence require a different type of response than what worked last time. We also consider the challenges of integrating host-based response systems and network-based systems.

Even when the threat has been neutralized or is otherwise no longer active, AICA must attempt to return the system to adequate working condition. In the "Recovery Planning" chapter, we introduce and demonstrate a recovery planning system that evaluates the impact of system degradation and generates COAs for recovery. The system evaluates these COAs through integrated heterogeneous simulations that account for unavoidable uncertainty and formally verify recovery COAs with confidence guarantees.

Given that hostile malware will give high priority to finding and disabling AICA, AICA must stay as undetectable as possible. In "Cyber Camouflage," we review the common techniques of adversarial reconnaissance, and methods for formally modeling reconnaissance activities and belief formation. We discuss common techniques such as honeypots, deceptive or obfuscated traffic, and deceptive responses to probes, and then propose new techniques based on adversarial machine learning to create more effective deceptive objects.

The "Adaptivity and Antifragility" chapter stresses the need to make cyber-defense agents adaptive and antifragile. A resilient system can survive attacks by autonomously adapting and managing its own functionality. An antifragile system can also enhance its capabilities and become more resilient as a result of endogenous and exogenous stressors. We present a concrete example of a self-improving system and middleware framework for antifragility.

When conditions permit, AICA may collaborate with other friendly agents. Conflicts may arise due to incompatible plans and objectives of the agents. Negotiations to jointly identify and execute a COA require building consensus under distributed and/or decentralized multi-agent settings with information uncertainties. The "Negotiation and Collaboration" chapter presents algorithmic approaches for enabling the collaboration and negotiation function. The strengths and limitations of potential techniques are identified, and a representative example is illustrated.

Humans are a special type of friendly agent, with special privileges. When conditions permit, human operators will oversee, approve or modify the actions of AICA. In the "Human Interactions" chapter, we explore knowledge acquisition to understand the user groups and use cases, as well as iterative design and feedback with users. Human trust in intelligent systems can be supported by transparency-based approaches, using metrics/frameworks to assess system transparency and explanation effectiveness.

The performance characteristics of AICA, including quality of defenses, resilience, reliability, probability of undesirable effects, etc., must be tested in measured in a consistent, quantitative and rigorous manner, under a broad range of conditions. The "Testing and Measurements" chapter draws upon real-world examples to present potential metrics for performance; reviews existing work in the field, e.g., several testbeds for testing autonomous cyber defense algorithms; and offers a detailed case study.

In "Deployment and Operation," we analyze several scenarios to consider the types of threats such agents might be expected to encounter and what actions would potentially be beneficial for them to take in response. These scenarios include an unmanned automated system (UAS, or "drone"), solo or as part of a swarm; an electrical distribution grid; an orbital or deep-space communication network; and a large-scale computational array (such as offered by a cloud vendor or high-performance computing).

Operations that involve a significant number of AICA-like agents comprise complex, intractable and risk-laden tasks. The "Command in AICA-intensive Operations" chapter explores how such operations would be commanded. A central part of managing these challenges is recognizing and accepting complexity. Additionally, success in AICA-intensive operations requires highly capable SU. Finally, the turbulent environment in which these units operate stresses the need for organizational agility, ensuring internal operations match the degree of turmoil in external environments.

AICA is essentially a robot that, by necessity, must be given a chain saw. As such, it presents a host of risks. In the "Risk Management" chapter, we argue that human intervention in real time during AICA operation may increase the harm. We propose other options for human-centered strategies that allow humans to shape AICA behavior before it chooses a COA: providing labeled data for supervised learning of AICA, offering a choice of machine learning algorithms, and/or devising algorithmic rules that constrain AICA actions.

Active autonomous systems like AICA face a host of policy issues, ethical concerns, governance concerns, societal impact concerns and legal concerns. In the "Policy Issues" chapter, we explore how the ever-changing concepts of cyber-defense reflect changing policies and review existing policies, including wartime policy considerations with examples. We note that, in some cases,

AICA may fall into the "gray zone" of policy and explore how this relates to US national and economic security policy, consumer privacy and matters of constitutional protections.

In "Development Challenges," we divide development challenges into two areas: engineering and research. The engineering ecosystem has six components: design, implementation, individual test and certification, composition, composite test and certification, and deployment. The research ecosystem includes models, architectures, mechanisms, testing and certification, operations, and social, ethical, and legal aspects. We show connections between these two ecosystems by describing how tackling challenges in the research ecosystem would contribute to tackling the challenges encountered when engineering AICAs.

The "Case Study A: The AICAproto21 Prototype" chapter describes a prototype system that encompasses a number of AICA features. This prototype was built using open-source software components in a containerized manner to allow for a quick time-to-completion with maximum flexibility for future capabilities. It demonstrated the ability of the agent to respond to an indicated attack with a defensive action. The chosen approach provided an easy-to-scale solution that is likely to work well cross-platform.

In chapter, "Case Study B: Tactical Edge Agent," we focus on aspects characteristic of deploying agents at the "tactical edge." Here the environment for an AI cyber-defense agent is vastly different from its classic habitat, the enterprise-scale network. We discuss our approach to overcoming the challenges of austere conditions, low availability of computing power, poor to nonexistent connectivity to enterprise-scale resources, and porous borders between the cyber domain (as conventionally considered) and the physical and electronic warfare (EW) domains.

A different type of resilience-support agents called sentinels are described in the "Case Study C: the Sentinel Agents" chapter. A sentinel agent is connected to the system interfaces from which it receives the data to support its monitoring function. The sentinel then conditions the diverse sets of collected data so that they can be integrated and analyzed, and performs the specific analyses required for detecting a cyber attack and determining the location within the protected system that is under attack.

## 8. Summary and Conclusions

AICA is an agent, i.e., a software that resides and operates on one or more computing devices, perceives and comprehends its environment, and plans and executes purposeful actions on the environment (and on itself) to achieve the agent's goals. AICA is local to a system and is responsible for defending the system from cyber compromises. If a compromise occurs, the agent is responsible for response and recovery of the system, usually autonomously. The autonomy of the agent is a necessity because of the growing scarcity of human cyber-experts who could defend systems, either remotely or onsite, and because sophisticated malware may degrade or spoof the communications of the system using a remote monitoring center. The agent can be distributed over multiple processes or devices, or could be implemented as a team of agents or subagents.

AICA observes the state and activities within the system it is asked to defend, diagnoses the situation and projects the future state of the system. AICA creates a plan of action, assesses the risks and benefits involved in the plans of actions and makes its decisions accordingly. Because AICA is responsible self-defense and self-preservation, it must practice stealth, camouflage and concealment. An AICA Reference Architecture has been proposed and defines five main functions: sensing and world state identification, planning and action selection, collaboration and negotiation, action execution and learning and knowledge improvement.

As AICA is intended to make changes in its environment, there is a risk that the agent's action could harm a friendly computer. This risk must be balanced against the losses that might occur if the agent does not act. To manage the risk, when able to communicate with the agent, the human supervisor could modify the goals, criteria and constraints of AICA or offer additional examples to improve the agent's learning process.

New technologies are often perceived as being associated with risk, thus human users have difficulties trusting the artifact. AICA's designers can help the trust-building process by providing the agent with a means to communicate to the users what the agent is doing, the decision process involved, and the information it used as inputs into its decision and eventual actions.

## References

Kott, A. and Arnold, C., 2013. The promises and challenges of continuous monitoring and risk scoring. IEEE Security & Privacy, 11(1), pp. 90-93.

Kott, A. and Linkov, I., eds., 2019. Cyber resilience of systems and networks. Springer International Publishing.

Kott, A. and Linkov, I., 2021. To improve cyber resilience, measure it. IEEE Computer, 54(2), Feb. 2021, pp. 80-85.

Kott, A. and McEneaney, W.M., 2006. Adversarial reasoning: computational approaches to reading the opponent's mind. Chapman and Hall/CRC.

Kott, A. and Stump, E., 2019. Intelligent autonomous things on the battlefield. In Artificial intelligence for the internet of everything (pp. 47-65). Academic Press.

Kott, A. and Théron, P., 2020. Doers, not watchers: intelligent autonomous agents are a path to cyber resilience. IEEE Security & Privacy, 18(3), pp. 62-66.

Kott, A., Golan, M.S., Trump, B.D. and Linkov, I., 2021. Cyber resilience: by design or by intervention?. Computer, 54(8), pp. 112-117.

Kott, A., Ludwig, J. and Lange, M., 2017. Assessing mission impact of cyberattacks: toward a model-driven paradigm. IEEE Security & Privacy, 15(5), pp. 65-74.

Kott, A., Théron, P., Drašar, M., Dushku, E., LeBlanc, B., Losiewicz, P., Guarino, A., Mancini, L., Panico, A., Pihelgas, M. and Rzadca, K., 2018. Autonomous intelligent cyber-defense agent (AICA) reference architecture. release 2.0. arXiv preprint arXiv:1803.10664.

Ligo, A.K., Kott, A. and Linkov, I., 2021. Autonomous cyberdefense introduces risk: can we manage the risk?. Computer, 54(10), pp. 106-110.

Ligo, A.K., Kott, A. and Linkov, I., 2021. How to measure cyber-resilience of a system with autonomous agents: approaches and challenges. IEEE Engineering Management Review, 49(2), pp. 89-97.

Linkov, I., Galaitsi, S., Trump, B.D., Keisler, J.M. and Kott, A., 2020. Cybertrust: From explainable to actionable and interpretable artificial intelligence. Computer, 53(9), pp. 91-96.

Muttik, I., 2016, Good viruses. Evaluating the risks. Talk at DEFCON-2016 Conference. https://www.defcon.org/images/defcon-16/dc16-presentations/defcon-16-muttik.pdf.

Théron, P., Kott, A., Drašar, M., Rzadca, K., LeBlanc, B., Pihelgas, M., Mancini, L. and de Gaspari, F., 2020. Reference architecture of an autonomous agent for cyber defense of complex military systems. In Adaptive Autonomous Secure Cyber Systems (pp. 1-21). Springer, Cham.

Théron, P., Kott, A., Drašar, M., Rzadca, K., LeBlanc, B., Pihelgas, M., Mancini, L. and Panico, A., 2018, May. Towards an active, autonomous and intelligent cyber defense of military systems: the NATO AICA reference architecture. In 2018 International conference on military communications and information systems (ICMCIS) (pp. 1-9). IEEE.